\documentstyle[11pt,epsf]{article}
\epsfverbosetrue
\newcommand{\bra}[1]{{\langle #1 \sbar}}
\newcommand{\ket}[1]{{\sbar #1 \rangle}}
\def\ket #1{|#1\rangle}
\def\bra #1{\langle #1 |}

\begin{document}

% title, authors and affiliation
\begin{center}
{\Large\bf Curvilinear coordinates for full-core atoms} \\
\vspace{0.5cm}
{A. Putrino and G. B. Bachelet} \\
\vspace{0.6cm} {\it Istituto Nazionale di Fisica della Materia (INFM) 
and Dipartimento di Fisica}\\
{\it Universit\`a di Roma La Sapienza, Piazzale Aldo Moro 5, I-00185 
Rome, Italy}\\
\end{center}

% abstract
\begin{abstract}
Curvilinear coordinates, introduced by F. Gygi for valence-only 
solids and molecules within the local-density functional theory, can be 
used to describe both core and valence states in 
electronic-structure calculations.  A simple and quite general 
coordinate transformation results in a large, yet affordable 
plane-wave energy cutoff for full-core systems (e.g., $\simeq 120$ Ryd 
for carbon or silicon) within the local-density functional theory, and 
in a reduced correlation time for full-core variational Monte Carlo 
calculations.  Numerical tests for isolated Li, C, and Si atoms are 
presented.
\end{abstract}
\vspace{0.2cm}

\section{\label{Introduction}Introduction}

The study of electrons in atoms, molecules and solids represents a 
very active front of computational and theoretical physics, where the 
frontiers of quantum statistical mechanics and materials science 
blend.  In the last 15 years at least two major innovations emerged in 
the context of large-scale electronic-structure calculations.
One is the Car-Parrinello method \cite{CRP85}, by which, within the 
Local Density Approximation (LDA) of the Density Functional Theory 
(DFT), the simultaneous time evolution of ionic positions and 
electronic wavefunctions became possible for realistic molecular or 
solid-state systems; in a sense, this is the most mature outcome of 
traditional self-consistent-field approaches.
Another is the new family of quantum Monte Carlo (QMC) methods, which 
use random numbers to sample realistic many-electron wavefunctions; 
their stage of development is less mature (the ionic motion is 
still out of the question), but they have already explored the 
electronic states of a number of atoms, molecules and solids, and 
appear as a promising tool for the future \cite{QMCRW}.

%accapo
Although their theoretical background and the corresponding 
computational strategies are quite different, both of these 
innovations have until now heavily relied on some pseudopotential 
approximation of the electron-ion interactions.
Real electrons in a coulombic nuclear potential are arranged according 
to the shell structure typical of atoms, and, unless the physically 
relevant valence information is extracted by some approximate 
valence-only scheme, most of the computational effort of either DFT or 
QMC methods is wasted in the great multiplicity of length and energy 
scales of inner, or core shells.
From this point of view the pseudopotential approximation 
\cite{xNCPP} usually does an excellent job, as witnessed by its 
extensive application to molecular and solid-state problems, but 
full-core approaches \cite{OKA75}, whenever possible, remain 
attractive, both for their ability to avoid ambiguities in those cases 
where the core-valence separation is less obvious, and, as we recall 
in a moment, for the less firm theoretical basis of pseudopotentials 
within the new Monte Carlo schemes.

%accapo
In this paper we propose a new method, based on curvilinear 
coordinates (hereafter CC), by which the valence properties of 
full-core atoms, rather than pseudo atoms, can be efficiently 
described in connection with both density functionals and 
variational Monte Carlo schemes.
In the remainder of this introduction we give a brief overview on the 
core-valence problem, which should highlight the background and 
motivations of our method, presented in Section \ref{Method}.  Section 
\ref{Results} will be devoted to numerical results on isolated atoms, 
and Section \ref{Conclusions} will present some conclusions and 
perspectives.

\subsection{\label{Full-core}{\it Full-core methods within the LDA}}

For isolated atoms, in spite of the many different length scales 
induced by the electronic shell structure as the atomic number $Z$ 
grows from 1 to 100, the use of radial logarithmic grids (with 
exponentially variable step, very dense near the nucleus, less and 
less dense towards the farther regions) allows an extremely accurate 
description of any atom in the Periodic Table of the Elements with 
just a few ($\simeq 200$) radial points, sometimes much less 
\cite{LWC65}.
After the advent of energy-linearized schemes \cite{OKA75}, this grid 
became a useful and effective tool also for solids or molecules, in 
connection either with atomic-sphere approximations (originally 
intended for closed-packed systems), or with a convenient partition of 
space (atomic spheres and interstitial regions).
In this case, however, the basis functions are quite different from 
plane waves or energy-independent atomic orbitals: the application of 
linear muffin-tin orbitals (LMTO) to open systems, like tetrahedrally 
bonded semiconductors or surfaces, required considerable theoretical 
developments \cite{MTH97}, and, at least up to now, even the powerful
linear augmented plane waves (LAPW) \cite{xLAPW} have not been used for 
Car-Parrinello-like molecular dynamics schemes.

\subsection{\label{Pseudopotentials}{\it Pseudopotentials}}

Smooth valence-only potentials which effectively replace the core 
electrons and the sharp nuclear coulomb singularity which binds them, or 
pseudopotentials, have always provided an attractive alternative.  In 
the late seventies and early eighties they acquired, within the DFT, 
explicit relations to the full-core atom, and became 
``first-principles'' pseudopotentials \cite{xNCPP}.
The price to pay for such a major upgrade was their angular 
nonlocality, which mainly reflects the different orthogonality effects 
felt by valence electrons of different angular momentum; this price 
was by far outweighted by the new ability of pseudopotentials: to 
predict equilibrium geometries, structural energies \cite{COH80}, and 
more recently extremely accurate phonon spectra \cite{BAR87} of 
valence-only solids and molecules with a relatively small plane-wave 
basis set.
In 1985 they thus provided a natural environment for the 
implementation of the powerful Car-Parrinello theory, by which the 
simultaneous time evolution of both ions and electronic wavefunctions 
became possible \cite{CRP85}.

%accapo
Electron-ion pseudopotentials turned out to be a necessary ingredient 
also for (more recent) stochastic approaches to the exact 
many-electron hamiltonian and ground-state wavefunction, like 
variational or diffusion Monte Carlo: even for QMC, up to now, the 
complication of a nonlocal valence-only ion has been by far preferable 
over the multiplicity of length and energy scales intrinsic to 
full-core atoms \cite{QMCPP}.
  
\subsection{\label{WhoCares}{\it Who cares about cores?}}

Approximate pseudoatoms are much simpler than true atoms, and even so 
pseudopotential simulations of large-scale molecular or solid-state 
systems remain a formidable computational task.
Standard first-principles pseudopotentials need as many as 200 plane 
waves per atom for a really accurate description of simple 
$s\!\!-\!\!p$ systems (i.e., a 20 Ryd energy cutoff in the case of, 
say, silicon), and thus many more for ``tough'' (but unfortunately 
very interesting) elements like carbon, oxygen, or first-row 
transition metals.
Even if softer pseudopotentials (e.g.  of the Vanderbilt type 
\cite{VDB90}) are adopted, Car-Parrinello simulations of hundreds of 
pseudoatoms require huge amounts of computer memory and time, and are 
seldom fully converged.
For a given system size, full-core approaches within the DFT (like 
LMTO or LAPW \cite{OKA75,MTH97}) end up with a computational effort 
which is not too different from a pseudopotential calculation, but, as 
mentioned, they cannot presently compete with plane waves as far as 
moving ions are concerned \cite{xLAPW}.
To find an alternative to pseudopotentials and describe, within the 
DFT, the molecular dynamics of full-core systems at a similar cost as 
pseudo systems (thus getting rid of pseudopotential approximations), 
represents a remarkable goal.

%accapo
More good reasons to seek new ways of describing full-core 
systems can be found outside the context of DFT.
Key aspects of the electron correlation in many interesting materials 
are not adequately described by approximate density functionals, and 
QMC methods are becoming a viable tool to study electron correlations 
exactly.  For this new family of stochastic many-body approaches 
isolated atoms (unlike LDA, where, using a radial logarithmic grid, 
they are a matter of seconds on any PC) turn out to be as difficult as 
polyatomic systems for $Z>10$ \cite{CEP86}.
A consistent construction of pseudopotentials from isolated atoms is 
therefore impossible (except for very light atoms \cite{ACI94}), and 
ionic pseudopotentials are thus normally obtained from LDA atoms and 
later used for QMC simulations \cite{QMCPP}.  This 
strategy is tenable for simple $s\!-\!p$ systems \cite{PPTHE}, 
but not for transition metals and other very interesting elements; its 
main justification is really the lack of any viable alternative.

%accapo
In conclusion, new methods aimed at full-core 
systems are both a desirable complement to existing DFT schemes and a 
necessary step forward in modern QMC methods.

\section{\label{Method} Method}

\subsection{\label{Background}{\it Background}}

Recently, in the context of DFT and pseudopotentials, F. Gygi 
\cite{GYG92} has shown that nonlinear coordinate transformations can 
provide a very efficient tool to reduce the size of the plane-wave set 
needed to describe a polyatomic system without reducing the accuracy 
of the calculation, or, in other words, that a given number of plane 
waves (equivalently, of spatial grid points) can describe a polyatomic 
system more accurately with CC than with regular euclidean 
coordinates.\footnote{A similar ``local scaling'' transformation was 
independently proposed for isolated atoms by Lude{\~n}a and coworkers 
with the rather 
different scope of constructing orbital-free energy-density 
functionals \cite{LUD91}.}
The new coordinates must become ``denser'' where the wavefunctions 
vary more rapidly (e.g.  near atoms and in the bond region for a 
covalent system), and less dense elsewhere (the interstitial regions).  
This can be efficiently accomplished by energy minimization (adaptive 
coordinates), as originally proposed by Gygi, but also, as he later 
showed \cite{GYG96}, by hanging to each atom of the molecule or solid 
under consideration an appropriate spherical deformation of the 
euclidean coordinates.

%accapo
Gygi's CC have been successfully applied by other groups; the 
applications have concerned pseudopotentials, valence-only systems and 
density functionals \cite{DRH96,LDS97}.  In this work we extend Gygi's 
CC to full-core systems.  Such a possibility was mentioned some time 
ago \cite{JOA93} and a very recent attempt in this direction has just 
been published \cite{MZK97}, but, as we will try to explain shortly, 
it bears little resemblance to ours.
Besides the advantages of CC for DFT studies of full-core systems, we 
propose here QMC simulations of full-core electronic systems as a new 
possible field of application; this idea was stimulated by a recent 
remarkable contribution of C. Umrigar, not based on CC \cite{UMR93}.

\subsection{\label{BasicIdea}{\it Basic idea}}

Three elements are at the origin of our work: Gygi's new method, a 
reconsideration of full-core approaches based on augmentation, and the 
idea that CC may become a general tool to study true atoms (made of 
core and valence electrons) even outside context of DFT, for example
within QMC methods.

%accapo
The partition of space in terms of atomic spheres and interstitial 
regions, typical of e.g.  LAPW, can be revisited in terms of Gygi's 
new concepts.  In the interstitial regions the electronic 
wavefunctions are quite smooth, and the uniform, low spatial 
resolution provided by a few plane waves is adequate.  In the atomic 
spheres, where core and valence wavefunctions undergo rapid spatial 
variations, the plane waves are augmented by atomic functions given on 
a radial logarithmic grid; such a high spatial resolution is 
equivalent to a strong spherical deformation of the euclidean space 
around each nucleus.
However in these methods the different spatial resolution adopted for 
different regions of space is not defined through a single, 
continuous, non-euclidean metric tensor; it instead undergoes a sharp 
discontinuity at the boundaries, which makes these methods radically 
different (and more involved) than those based on a fixed basis set, 
like ordinary plane waves.

%accapo
In this work we propose (and numerically test for isolated atoms) the 
idea that a suitable, continuous, non-euclidean metric tensor, can 
account for the strong inhomogeneities of a real electronic system, 
or, in other words, that $-Z/r$ nuclear potentials, core electrons, 
and valence electrons, can all be efficiently represented on a 
suitable set of spatially continuous, energy-independent basis 
functions of the Gygi type.  Besides DFT applications, we will 
try to show that such a CC transformation may be a promising 
tool for QMC methods.

\subsection{\label{RadialTransf}{\it The radial transformation}}

The key ingredient of our method is a coordinate transformation which 
deforms the euclidean space around each atom in such a way as to mimic 
a logarithmic grid in the core region and to smoothly merge into a 
uniform grid away from it.
Our transformation has the same goal as that independently proposed by 
Modine {\it et al.} \cite{MZK97}, but it's based on a quite different 
strategy, and, in our intentions, it should be much simpler and more 
physically transparent.  Their rather complicated two-step procedure 
breaks many symmetries and is neither a true adaptation, nor a 
coordinate transformation which explicitly follows the ionic 
positions.
Ours preserves all crystal symmetries, being based on the 
superposition of spherical deformations centered on the atoms.  Around 
each atom we propose a strong radial deformation $\nu(\rho)$, whose 
range $\rho_{o}$ approximately corresponds to the covalent radius:

\begin{equation}
r=f(\rho) = C\!\rho\, {\rm exp}\left(b\rho\right)\,\exp\left[-\left({\rho\over
\rho_{o}}\right)^{4}\right]+\rho\,\left\{1-
\exp\bigg[-\left ({\rho\over
\rho_{o}}\right )^{4}\bigg] \right\}
=\left[1+\nu\left(\rho\right)\right]\rho
\label{transformation}
\end{equation}

\begin{figure}
\epsfclipon
\epsfysize=12cm
\centerline{\epsffile{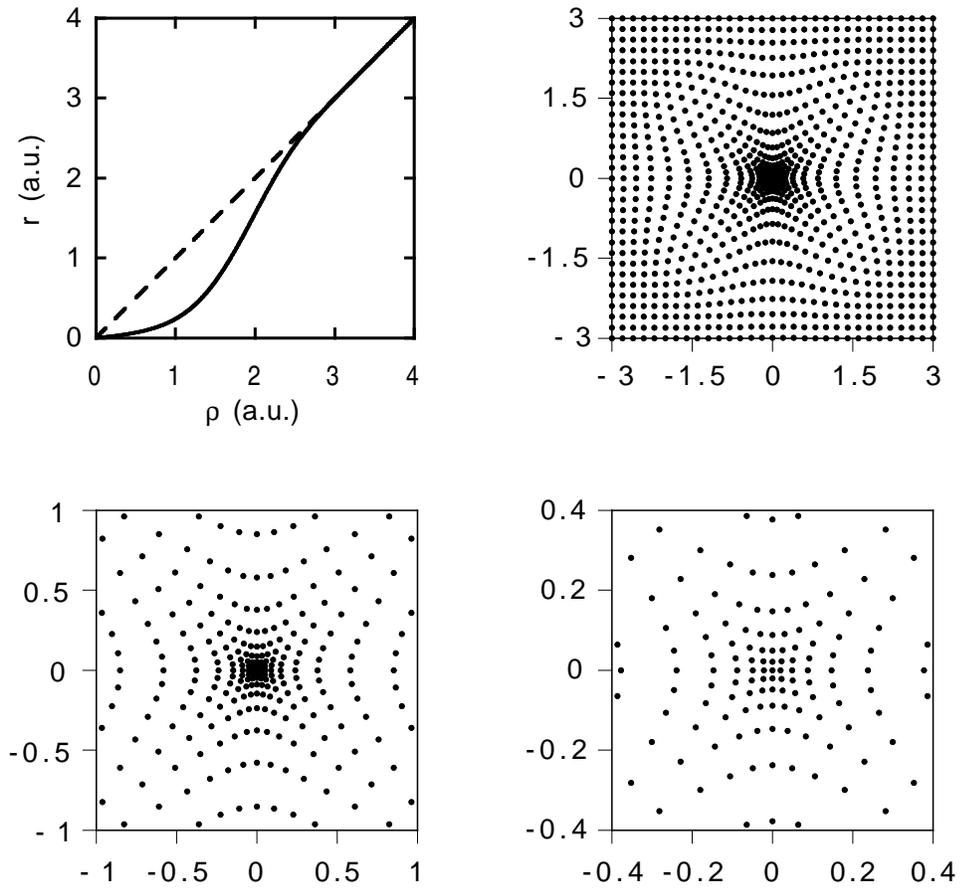}}
\caption{\protect\small Coordinate transformation for atomic silicon.  
Upper left panel: radial transformation $r=f(\rho)$ 
(Eq.~\ref{transformation}, solid curve); $r=\rho$ is shown as a dashed 
line.  Upper right panel and lower panels: a $2D$ uniform grid 
$\vec\rho$ maps into a real-space $2D$ nonuniform grid $\vec r$ 
(dots).  We need three different length scales to appreciate the 
effect of our transformation (see text).}
\label{fig1-grid}
\end{figure}

\begin{figure}
\epsfclipon
\epsfysize=5cm
\centerline{\epsffile{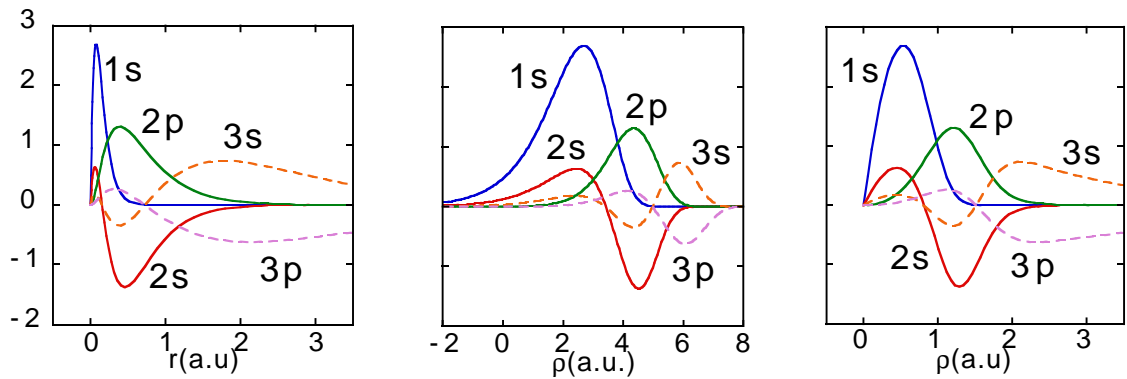}}
\caption{\protect\small Radial wavefunctions $u_{n \ell}=rR_{n \ell}$ 
for atomic silicon as a function of the distance from the nucleus.  
Left: ordinary radial coordinate $r$.  Center: logarithmic radial 
coordinate $\rho=ln(r/r_{0})/\alpha$ with $\alpha = 1$ a.u.  Right: 
radial coordinate $\rho=f^{-1}(r)$ (the inverse of our 
Eq.~\ref{transformation}, Fig.~\ref{fig1-grid}).  As a function of $r$ 
(left) the core and valence wavefunctions have a very different length 
scale, but on a logarithmic grid (center) or using our radial 
transformation (right) they acquire a similar length scale.}
\label{fig2-radwf}
\end{figure}

\noindent Here $\rho$ is the new radial coordinate, $r$ the actual 
distance from the nucleus, and $\nu$ the radial deformation with respect
to the euclidean case $r=\rho$.
The above transformation, optimized for 
the isolated silicon atom, is shown in the upper left panel of 
Fig.~\ref{fig1-grid} ; its purpose is to shift spatial resolution from 
the valence to the core region, so that core and valence shells 
acquire a similar length scale in the new radial coordinate (see 
Fig.~\ref{fig2-radwf}).

%
% figure 1 starts
% figure 1 ends
%

%
% figure 2 starts
% figure 2 ends
%

\noindent The transformation (Eq.~\ref{transformation}) was chosen to 
satisfy the following properties:

%accapo
\vbox{
\begin{description}
\item[(a)] unlike the logarithmic mesh, it maps $(0,\infty)$ into 
$(0,\infty)$, to allow the superposition of many such transformations 
centered at different sites, and thus a generalization to polyatomic 
systems;
\item[(b)] in the origin it recovers a regular, linear behavior $r 
= C\!\rho$;
\item[(c)] it goes back to ``flat-space'' $r=\rho$ outside the atom;
\item[(d)] it's logarithmic-like in between;
\item[(e)] it's invertible and differentiable.
\end{description}
}

%accapo
\noindent In a molecule or a crystal the spatial CC transformation is 
obtained as a superposition of spherical deformations $\nu(\rho)$ 
centered on the atoms:\footnote{Since we give the radial 
transformation as $r(\rho)$ and not as $\rho(r)$, 
Eq.~\ref{superposition} is more convenient than Gygi's definition 
\cite{GYG96}; they become identical only if the deformations relative 
to different atoms do not overlap.}

\begin{equation}
        {\vec r} = 
        {\vec\rho} +
        \sum_{\alpha} (\vec\rho - {\vec R}_{\alpha})
        \nu_{\alpha}(\mid\vec\rho - {\vec R}_{\alpha}\mid)
        \label{superposition}
\end{equation}

%accapo
\noindent In the upper right panel and in the two lower panels of 
Fig.~\ref{fig1-grid} we show how a uniformly spaced two-dimensional 
grid in the new coordinates $\vec\rho=(\xi,\eta)$ maps into a 
nonuniform grid of real-space points $\vec r=(x,y)$ (shown as dots) 
for an isolated silicon atom, placed in the origin.
In the upper right panel $(x,y)\in (-3,3)$ a.u.  Near the borders of 
the square, for $r >2$, our transformation recovers the identity $\vec 
r = \vec\rho$, and a uniform $\vec\rho$ grid maps into a uniform $\vec 
r=\vec\rho$ grid.  Towards the atom the $\vec r$ grid points become 
denser, and cannot be resolved by eye.

%accapo
We then try to zoom in, but even in the lower right panel, $(x,y)\in 
(-1,1)$ a.u., we cannot resolve the $\vec r$ grid points of the inner 
region: between $r=0.2$ and $r=2$ the transformation mimics a 
logarithmic grid, and the more we approach the nucleus, the higher is 
the density of $\vec r$ points.
By further zooming in, for $(x,y)\in (-0.4,0.4)$ a.u.  (lower right 
panel), our eye finally appreciates, below the $1s$ radius (compare to 
Fig.~\ref{fig2-radwf}), the recovery of a uniform grid, as implied by 
the properties (a) and (b) of Eq.~\ref{transformation}: very close to 
the nucleus our transformation recovers a linear behavior $\vec r = 
C\!\rho$ (with $0 < C \ll 1$).

%accapo
Fig.~\ref{fig2-radwf} shows that the effect of our transformation 
(right panel) is, as desired, similar to that of a logarithmic grid 
(center panel); the two main differences are related to the properties 
(a-c), i.e., the behavior near the origin and the recovery of a 
uniform euclidean resolution beyond the outermost peak of the valence 
shells.  Figs.~\ref{fig1-grid} and \ref{fig2-radwf} refer, as said, to 
the silicon atom.  Of the three parameters which characterize the 
transformation Eq.~\ref{transformation} one, $\rho_{o}$, it's 
approximately equal to the covalent radius because of the property 
(c), and thus practically fixed; the other two, $C$ and $b$, are 
positive parameters which are optimized for each particular atom to 
``equalize'' core and valence resolution.  Their values, for silicon 
as well as the other atoms considered in this work, are given in 
Table~\ref{tab1-params}.

\begin{figure}
\epsfclipon
\epsfysize=6cm
\centerline{\epsffile{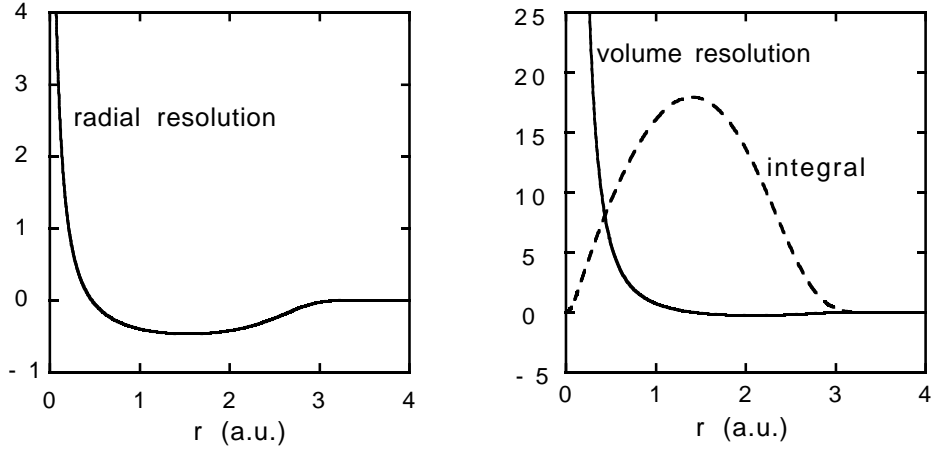}}

\caption{\protect\small Left: $({d\rho/dr})-1$, difference between the 
radial resolution $({d\rho/dr})$ of the new CC 
(Eq.~\ref{transformation} and Fig.~\ref{fig1-grid}) and $1$, the 
uniform radial resolution of ordinary euclidean coordinates.  Right, 
solid line: $(\rho^{2}d\rho/r^{2}dr) -1$, difference between the 
volume resolution of the new CC and the uniform volume resolution of 
ordinary euclidean coordinates.  Right, dashed line: volume integral 
of $(\rho^{2}d\rho/r^{2}dr) -1$ (see text).}
\label{fig3-density}
\end{figure}

%accapo
To equalize core and valence resolution amounts to increasing the 
spatial resolution in the core region and to decreasing it in the 
outer valence region in such a way as to recover a uniform resolution 
outside the atom.  The effect is illustrated for the silicon atom in 
Fig.~\ref{fig3-density}, where two key differential properties of our 
radial transformation are shown.
%
% figure 3 starts
% figure 3 ends
%
In the left panel we subtract the unity from the radial resolution 
corresponding to our transformation (Eq.~\ref{transformation} and 
Fig.~\ref{fig1-grid}).  In this way we visualize the comparison 
between our new radial resolution $d\rho/dr$ and the uniform, unit 
radial resolution of the identical transformation $\rho=r$ (i.e., of 
ordinary euclidean coordinates).
Where $d\rho/dr -1$ is positive we are gaining radial resolution; 
where it's negative we are losing radial resolution; where it's zero 
(outside the covalent radius) we recover the unit resolution of 
ordinary euclidean coordinates.

%accapo
The same comparison is shown in the right panel, but now for the {\it 
volume} resolution.  Here (solid line) we subtract the unity from our 
volume resolution $\rho^{2}d\rho/r^{2}dr$.

Again, where $(\rho^{2}d\rho/r^{2}dr) -1$ is positive, in the inner 
atomic region, we are gaining volume resolution; where it's negative, 
in the outer valence region, we are losing volume resolution; where 
it's zero, outside the covalent radius, we have the same resolution as 
ordinary euclidean coordinates.

%accapo
Our new CC are given in terms of the radial transformation 
Eq.~\ref{transformation}, and the radial resolution (left panel) is 
certainly interesting, but the volume resolution (right panel, solid 
line), which is the jacobian determinant of the spatial transformation 
(\cite{GYG92,GYG96}), is in fact the right thing to look at when 
comparing the local resolution of the new coordinate system against 
euclidean coordinates.
Let us thus concentrate on the right panel of Fig.~\ref{fig3-density}.  
Below $r=1.5$ a.u.  there is a large gain in the volume resolution (a 
larger and larger gain as the nucleus is approached: here the highest 
resolution is needed, because both core and valence wavefunctions 
oscillate rapidly); this gain is compensated by a slight loss above 
$r=1.5$ a.u., where the valence wavefunctions have their outermost 
maximum.

%accapo
The compensation of a gain with a loss is inevitable when the 
deformation of the euclidean coordinates is confined within a finite 
volume $V$ and vanishes outside it: in our case the atomic volume, 
approximately defined by the covalent radius due the property (c).  
Such a compensation corresponds in fact to a simple conservation law, 
illustrated by the dashed curve in the right panel.
The dashed curve represents the volume integral of the volume 
resolution, and falls to nothing oustide the atom, telling that a 
finite volume of the $\vec \rho$ space (in our case a sphere of radius 
$r_{cov}$) maps into the same finite volume of the $\vec r$ space (a 
sphere of radius $\rho_{cov}=r_{cov}$).

\begin{figure}
\epsfclipon
\epsfysize=6cm
\centerline{\epsffile{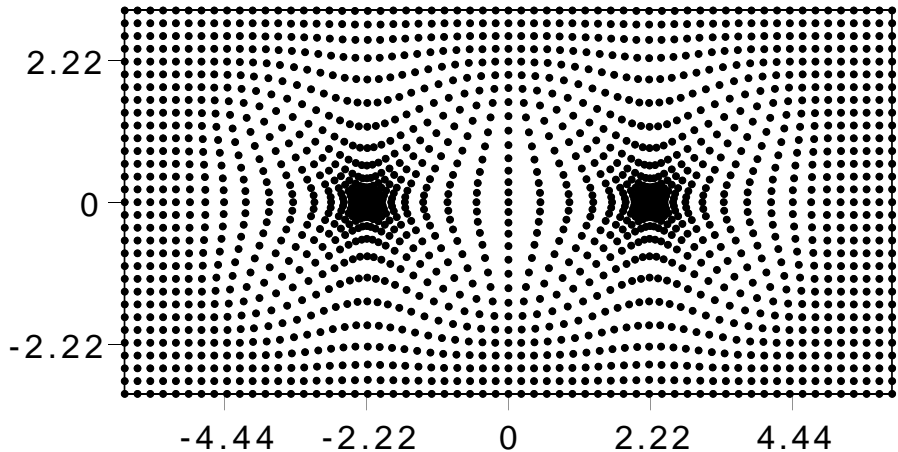}}
\caption{\protect\small Coordinate transformation for a silicon dimer.  
A $2D$ uniform grid 
$\vec\rho$ with $\Delta\rho_{x}=\Delta\rho_{y}=0.2$ a.u.  maps into a 
real-space $2D$ nonuniform grid $\vec r$ (dots). Blowing up the inner-core
region (black spots) yields identical results as the isolated atom 
(Fig.~\ref{fig1-grid}).}
\label{fig4-dimer}
\end{figure}

%accapo
The condition (c) could in principle be relaxed to allow 
$r=\gamma\rho$ with $\gamma > 0$ but different from unity outside the 
atomic volume $V$; in this case the ordinary atomic volume $V={4 \over 
3}\pi r_{cov}^{3}$ maps into a different CC atomic volume 
$V^{\prime}={4 \over 3}\pi \rho_{cov}^{3}=V/\gamma^{3}$, and, in a 
crystal, lattice constants and reciprocal vectors for euclidean 
coordinates and CC would differ by a scale factor $\gamma$.  For 
$\gamma > 1$ this would amount to a net saving of some useless volume 
resolution in the interstitial regions, which at first sight could 
seem an advantage.  However, until the CC transformation is just a sum 
of spherical deformations centered in the atoms as in 
Eq.~\ref{superposition}, such a choice would reduce the volume 
resolution not only in the empty interstitial regions, but also in the 
midbond region, which is a great disadvantage; that's why we kept the 
property (c) as it is.
In fact (as illustrated by Fig.\ref{fig4-dimer} which shows the CC 
appropriate to a silicon dimer), even with $\gamma=1$, our 
superposition of atomic deformations suffers of some loss of 
resolution in the bond region.  This is inevitable, because we have 
optimized our transformation in the isolated atom, where we gain 
resolution in the core region at the expense of the valence region; 
for this reason, as we will see in the next section, full-core atoms 
will require an effort which is affordable and much smaller than with 
euclidean coordinates, but still larger than pseudo atoms.

%
% figure 4 starts
% figure 4 ends
%

From Fig.~\ref{fig4-dimer} we see that, to further improve the 
efficiency of our superposition of atomic deformations in a molecule 
or a solid, it would be good to shift some useless volume resolution 
away from the empty regions and bring it into the bond region.  For 
this aim a better idea than playing with $\gamma$ would probably be to 
add to the superposition of atomic deformations 
Eq.~\ref{superposition} some appropriate deformation centered at the 
interstitial sites or in the midbond.  This kind of upgrade can 
further improve the efficiency of the atomic results presented here, 
and must be tested and optimized in the molecule or solid.

%accapo
Here we limit ourselves to the atomic transformations, and in the 
following Sec.~\ref{Results} we will test them in two ways: 
generalized plane waves within LDA (\ref{LDA-PW}), and variational 
Monte Carlo (\ref{VMC-results}).

\section{\label{Results} Atomic results}

\subsection{\label{LDA-PW}{\it LDA: generalized plane waves}}

\begin{figure}
\epsfclipon
\epsfysize=6cm
\centerline{\epsffile{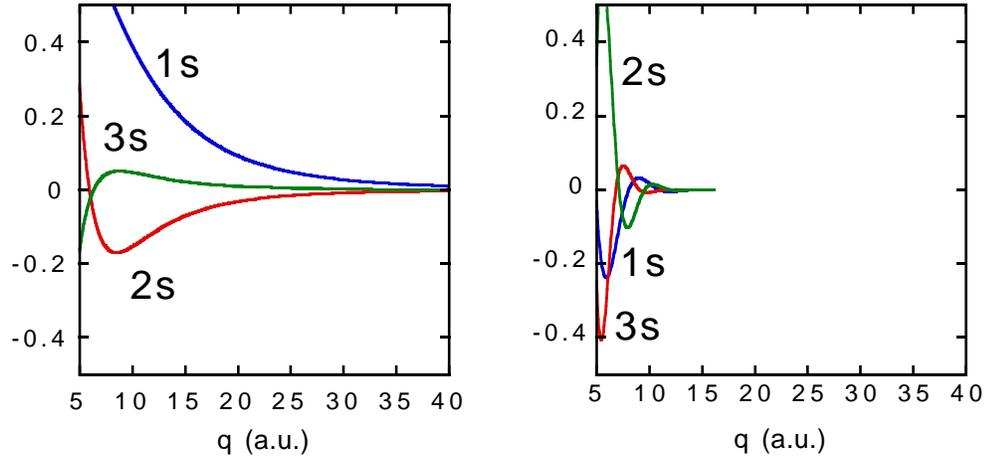}}
\caption{\protect\small Fourier transform (left) and Gygi transform 
(right) for the radial $s$ wavefunctions of an isolated atomic silicon.}
\label{fig5-fourier}
\end{figure}

With our new CC the core and valence radial wavefunctions acquire a 
similar length scale (Fig.~\ref{fig2-radwf}); as a consequence, their 
representation in terms of the corresponding Gygi's generalized plane 
waves \cite{GYG92,GYG96} converges for a much smaller wavevector $q$ 
than it would happen with regular plane waves and euclidean 
coordinates. This is clearly illustrated by Fig.~\ref{fig5-fourier}.
In the left panel we plot, as a function of the wavevector $q$, the 
ordinary Fourier transforms of the $1s$, $2s$, and $3s$ wavefunctions 
of the isolated silicon atom (obtained from a self-consistent LDA 
calculation).
In the right panel we plot the corresponding ``Gygi transforms'' on 
the same scale.\footnote{We only show $q \geq 5$ to emphasize the 
high-$q$ convergence.  For $q < 5$ (not shown) the absolute value of 
Fourier and Gygi transforms is similar, and much larger: for the 
$n,\ell$ shown in Fig.~\ref{fig5-fourier} they range from twice to 
$\sim 120$ times its full scale.}

%accapo
While the absolute value of the ordinary Fourier transforms (left) 
drops below a tenth of the full scale of Fig.~\ref{fig5-fourier} only 
for $q > 25$ a.u.  (corresponding to a plane-wave cutoff energy 
$E_{cut} \gg 625$ Ryd) and remains well above one hundredth of the 
full scale of Fig.~\ref{fig5-fourier} even at $q=40$ a.u.  
($E_{cut}=1600$ Ryd), the absolute value of the corresponding Gygi 
transforms drops below a tenth of the full scale of 
Fig.~\ref{fig5-fourier} already by $q=9$ a.u.  ($E_{cut}=81$ Ryd), and 
below one hundredth between $q=11$ and $q=12$ a.u.  ($E_{cut}=144$ 
Ryd); it becomes completely invisible, on the scale of our plot, for 
$q> 14$ a.u.  ($E_{cut}\simeq 200$ Ryd).

%
% figure 5 starts
% figure 5 ends
%

\begin{figure}
\epsfclipon
\epsfysize=6cm
\centerline{\epsffile{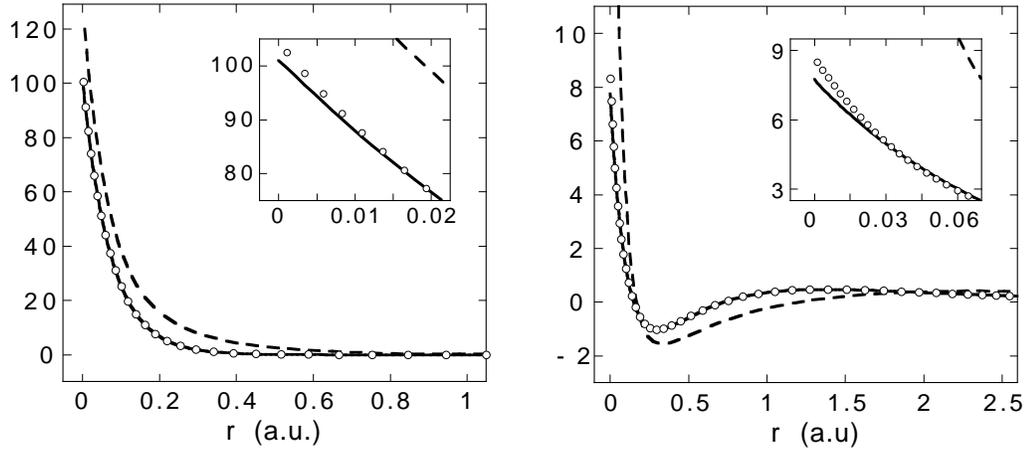}}
\caption{\protect\small Comparison of the fully converged
$1s$ (left) and $3s$ (right) radial wavefunctions of silicon, shown as 
solid lines, and the backtransformation of their Gygy transforms with
$E_{cut}=25$ Ryd (dashes) and $E_{cut}=100$ Ryd (empty dots). By
$E_{cut}=100$ Ryd the functions are perfect everywhere except very 
close to the origin (see insets and text).}
\label{fig6-backtrans}
\end{figure}

If we backtransform the right-hand curves of Fig.~\ref{fig5-fourier} 
up to a finite wavevector $q_{max}$ we thus expect a very good 
convergence for any $q_{max} \geq 9$ ($E_{cut} \geq 81$ Ryd).
This is shown in Fig.~\ref{fig6-backtrans}, where we compare the $1s$ 
(left panel) and $3s$ (right panel) original radial wavefunctions 
(solid line) with the backtransformation of their Gygi transforms with 
a low cutoff $q_{max}=5$ a.u.  (dashes) and a high cutoff $q_{max}=10$ 
a.u.  (empty dots).  We have chosen to show the core $1s$ because, as 
we will see in a moment, it's of crucial importance for the energy 
convergence, and the $3s$ as a typical valence radial wavefunction, 
but the same behavior as a function of $q_{max}$ is found for the 
occupied states of any other $n$ and $\ell$.
As expected from Fig.~\ref{fig5-fourier}, by $q_{max}=10$ a.u.  
($E_{cut}=100$ Ryd) the convergence is so good that we cannot 
distinguish the original radial wavefunctions from their 
backtransformed counterparts.

%
% figure 6 starts
% figure 6 ends
%

Besides the visual impact of Figs.~\ref{fig5-fourier} and 
\ref{fig6-backtrans}, it's a good idea to quantitatively check the 
convergence of the total electronic energy as a function of the cutoff 
$q_{max}$, or, equivalently, of the cutoff energy $E_{cut}$.  We do 
that by cutting off the Gygi transforms of all the occupied atomic 
wavefunctions (shown in Fig.~\ref{fig5-fourier} for $\ell = 0$) 
beyond different $q_{max}$ values, backtransforming them to real 
space, and monitoring the resulting total electronic energy as a 
function of $q_{max}$.\footnote{We do not perform a different 
self-consistent calculation for each $q_{max}$, but simply expand
the self-consistent wavefunctions of a converged real-space atomic 
calculation in terms of a finite number of Gygi spherical waves 
\cite{GYG92,GYG96}.}

\begin{figure}
\epsfclipon
\epsfysize=6cm
\centerline{\epsffile{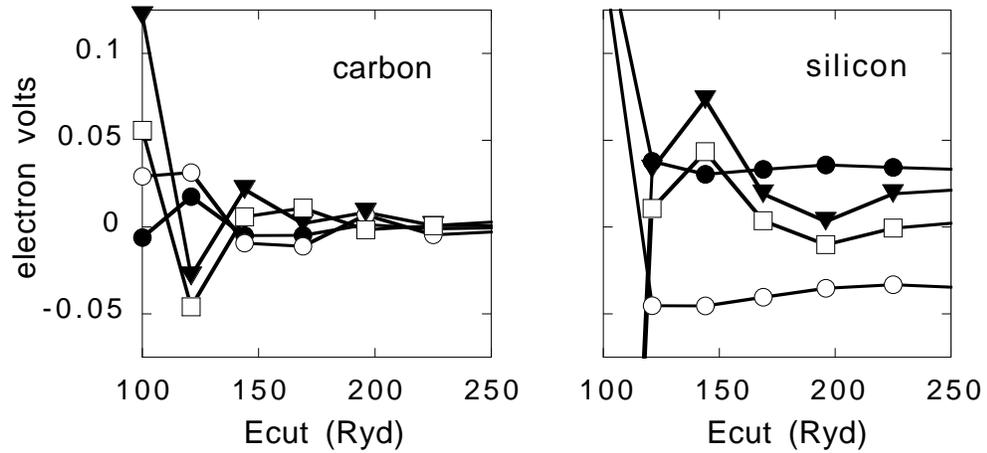}}
\caption{\protect\small Error on the ionization potentials (1st: full 
dots; 2nd: empty dots; 3rd: full triangles) and on the hybridization 
energy (empty squares) as a function of the cutoff.  Beyond $E_{cut} 
\simeq 120$ Ryd the errors on these valence energy differences are 
hundredths of an $eV$, although the two individual atomic energies 
still suffer from an error in the $eV$ range.  Left: carbon.  Right: 
silicon.}
\label{fig7-ionpots}
\end{figure}

%accapo
The results are shown in Table~\ref{tab2-ionpots} for atomic carbon 
and silicon; for atomic silicon they are also displayed in the left 
panel of Fig.~\ref{fig7-ionpots}.  The error, shown as a function of 
the cutoff energy (in Rydbergs), is everywhere defined as the energy 
deviation (in electron volts) from a perfectly converged real-space 
LDA atom.\footnote{Relativistic and spin-polarization effects are 
consistently not included in this work, so our results cannot be 
directly compared to experimental atomic energies.}

%accapo
Let us first concentrate on the second column of 
Table~\ref{tab2-ionpots}, which contains the total-energy error (in 
electron volts) for the neutral carbon and silicon atoms.
We immediately see that, even for the highest cutoffs, the error on 
absolute total energies is of the order of a few electron volts.
This error is in fact of the same order of magnitude as that found by 
Modine {\it et al.} on absolute total energies for isolated carbon and 
oxygen atoms, and could sound as a discouraging result: it tells us 
that a cutoff energy of $100-150$ Ryd, which corresponds to visually 
perfect backtransforms of the atomic wavefunctions 
(Fig.~\ref{fig6-backtrans}), yields an error which is certainly tiny 
on the scale of the core and total atomic energies (one to two parts 
per thousand), but unfortunately still sizable on the scale of the 
physically relevant valence energies.  We have tried to carefully 
analyze the origin of such an absolute total-energy error, since we 
suspected, on transparent physical grounds, that it cancels out in 
binding energies and other differential valence properties.
Our finding is that the error mostly comes from the $1s$ state, and is 
indeed almost independent of the valence configuration.  A simple 
mathematical argument shows that our transformation can greatly 
reduce, but not eliminate, the problem of the electron-nuclear 
cusp.\footnote{The cusp also affects the higher $s$ states, but in 
this case, for obvious reasons, its quantitative impact on total 
energies is practically negligible.}  Such an origin of the $eV$-sized 
error in the total energies is confirmed by monitoring the individual 
contributions to the total energy, with special attention to $\bra{n 
\ell}-{1\over 2} \nabla^{2}-Z/r\ket{n \ell}$ \cite{PUT97}.  As a 
result, valence properties can be obtained to a high accuracy with a 
large but affordable cutoff: for example a $\sim 120$ Ryd cutoff 
allows estimates of valence ionization potentials and hybridization
energies which are accurate within a few hundredths of an eV 
for Carbon and Silicon atoms (see Table \ref{tab2-ionpots} and 
Fig.~\ref{fig7-ionpots}).  Another way of saying that any total-energy 
contribution other than $\bra{n \ell}-{1\over 2}\nabla^{2}-Z/r\ket{n 
\ell}$ converges much faster than it, is to say that our CC are 
particularly effective for the solution of the electrostatic problem.

%
% table 1 starts
\begin{table}[tbp]

\begin{center}
\caption{\protect\small Parameters of the transformation 
(Eq.~\ref{transformation})}

\begin{tabular}{|c|c|c|c|}
        \hline
    & C & b & $\rho_{0}$  \\
        \hline
Li & 0.42-0.91 &  0.1-0.2 & 1-7  \\
        \hline
C & 0.139 &  0.75 & 1.75  \\
        \hline
Si & 0.089 & 0.75 & 2.0  \\
        \hline
U & 0.03 & 0.75 & 2.3  \\
        \hline
\end{tabular}

\label{tab1-params}
\end{center}
\end{table}
%table 1 ends
%
%
% table 2 starts
\begin{table}[tbp]

\begin{center}
\caption{\protect\small Energy errors$^*$ (in $eV$) versus cutoff energy 
$E_{cut}$ (in Rydbergs).}

\begin{tabular}{|c|c|c|c|c|c|c|}
\hline\hline
  & ${E_{cut}}$ & ${\rm Total}$ & ${\rm 1st \, IP}$ & 
  ${\rm 2nd \, IP}$  & ${\rm 3rd \, IP} $ & 
  ${\rm sp}^{{}_{\scriptstyle 3}}$\\
\hline
carbon &  $81$ & $8.1634$ &  $0.4299$ & $0.7626$ & $-0.9114$ & $-0.8881$\\
  & $100$ & $8.1090$ &  $0.0771$ & $0.1671$ & $-0.0743$ & $-0.2022$\\
  & $121$ & $3.4015$ & $-0.0016$ & $0.0039$ &  $0.0067$ & $-0.0091$\\
  & $144$ & $1.5238$ &  $0.0044$ & $0.0028$ & $-0.0498$ & $-0.0251$\\
\hline
silicon & $100$ & $22.340$ &   $0.172$ &  $0.154$ &  $-0.562$ &  $-0.587$\\
  & $121$ & $28.436$ &   $0.038$ & $-0.045$ &   $0.033$ &   $0.011$\\
  & $144$ & $15.701$ &   $0.030$ & $-0.045$ &   $0.073$ &   $0.043$\\
  & $169$ &  $5.932$ &   $0.033$ & $-0.040$ &   $0.019$ &   $0.004$\\
\hline\hline
\end{tabular}

\label{tab2-ionpots}

\footnotesize{$^*$Total: total energy; IP: 
ionization potential; ${\rm sp}^{{}_{\scriptstyle 3}}$: hybridization 
energy. Beyond $E_{cut}\!\simeq\! 100$ Ryd}\par
\footnotesize{the error on valence energies 
(last four columns), which are total-energy differences, is very 
small.}

\end{center}
\end{table}
%table 2 ends
%

%accapo
For example, by $E_{cut} \simeq 70$ Ryd for the silicon atom, and by 
$E_{cut} \simeq 170$ Ryd for a huge atom like uranium, the absolute 
electrostatic energies are within their converged value by less than 
$0.1 eV$; the uranium results compare thus very well with Goedecker's 
recent electrostatic investigations based on wavelets \cite{GOE97}.

%
% figure 7 starts
% figure 7 ends
%

It's of course very easy to test our CC for self-consistent Kohn-Sham 
atoms in real space, by simply generalizing the usual algorithm for 
logarithmic grids \cite{LWC65} to any radial grid.  Accurate 
ionization potentials and hybridization energies are obtained with 
much less grid points than total energies, but even absolute total 
energies converge quickly with the number of grid points (more quickly 
than they do with the maximum wavevector in the previous examples), 
because here the usual Taylor expansion near the origin takes care of 
the electron-nuclear cusp problem \cite{PUT97}.
This finding serves thus only as a final confirmation of our analysis 
of the total-energy error found with generalized plane waves: for 
truly three-dimensional real-space calculations (solids or supercell 
molecules) based on CC we don't have at hand an easy way to take 
advantage of the exact Taylor expansion near the nucleus, and thus we 
expect a similar performance as the generalized-plane-wave expansion 
of the previous subsection \ref{LDA-PW}: good valence energies in 
spite of fair total energies.

Although accurate valence properties represent already a remarkable 
achievement, it would be of course very desirable to have an efficient 
way of totally defeating the electron-nuclear cusp both in generalized 
plane waves and in three-dimensional real space.  An easy 
remedy,suggested by Modine {\it et al.}, is to replace the nuclear 
point charge with a gaussian charge whose range is much smaller than 
the $1s$ radius.  Unfortunately, by doing that, the resulting absolute 
total-energy error is still in the $eV$ range for atoms like carbon or 
silicon (i.e., of the same order of magnitude as doing nothing), as 
can be easily understood and numerically checked; however, in the lack 
of better ideas, such a remedy can become of interest when moving ions 
in a Car-Parrinello molecular-dynamics simulation.

\subsection{\label{VMC-results}{\it Variational Monte Carlo}}

%
% figure 8 starts
\begin{figure}
\epsfclipon
\epsfysize=7cm
\centerline{\epsffile{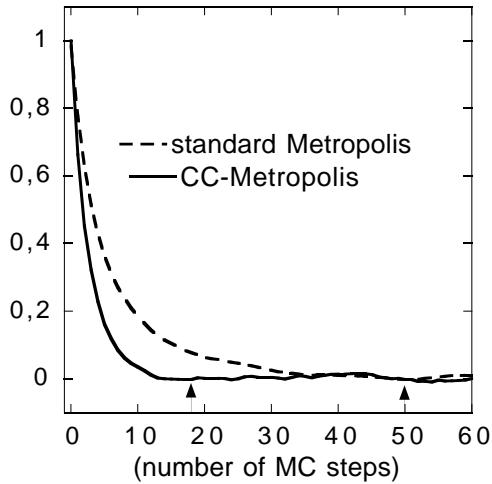}}
\caption{\protect\small Energy autocorrelation function for the 
standard Metropolis algorithm (dashed line) and for the CC Metropolis 
algorithm (solide line), when used for the lithium atom.  The two 
arrows approximately indicate the autocorrelation time in either case.}
\label{fig8-vmc}
\end{figure}
% figure 8 ends
%

In this section we show preliminary results for the isolated lithium 
atom which suggest that a straight Metropolis sampling in the new CC 
improves the sampling efficiency by a factor of three with respect to 
a straight Metropolis sampling in the ordinary euclidean coordinates.
This new application of CC was somewhat inspired by a recent 
contribution of C. Umrigar in the field of Variation Monte Carlo 
methods \cite{UMR93}.  He has proposed a generalized Metropolis scheme 
which, by a smart modification of the proposal matrix, obtains tiny 
correlation times and high acceptance rates, thus efficiently reducing 
the inconvenience of multiple length scales of electrons in real 
isolated atoms.  His (rather impressive) results concern both 
first-row elements, and the heavier argon atom ($Z=18$), but, unlike 
CC, cannot be easily generalized to polyatomic systems.
When plugged into the variational Monte Carlo scheme, a CC 
transformation modifies the proposal matrix with respect to
a standard Metropolis sampling and can be tailored to 
correspondingly reduce the autocorrelation time in a 
similar, but simpler way, as the Accelerated Metropolis algorithm of 
Umrigar \cite{UMR93}.
With respect to his approach our coordinate transformation shares a 
variable length for the proposed step as a function of the distance 
from the nucleus.  In the preliminary results presented here we don't 
take any further advantage of the wavefunction (as Umrigar, instead, 
does \cite{UMR93}); so our minimum autocorrelation time will still
correspond to an acceptance ratio of $\sim 50\%$, and we expect 
an interesting, but less 
pronounced improvement over straight Metropolis MC, than him.
In Fig~\ref{fig8-vmc} we show the energy autocorrelation function 
obtained from standard and CC Variational Monte Carlo runs for the 
lithium atom \cite{PUT97}.
We do find a reduced correlation time (by a factor of $\sim 3$).  
Similar results can be obtained with a relatively wide range of 
parameters (see Table~\ref{tab1-params}; the results of Fig~\ref{fig8-vmc}
correspond to the particular choice $C=0.42$, $b=0.2$, $\rho_{o}=7$).
As expected, this is good but less spectacular than Umrigar's results.  
The real advantage of our approach based on CC is that it lends itself 
to a simple and natural generalization to polyatomic systems.  Our VMC 
experiment in CC is the first step in an entirely new direction; many 
more are of course needed to explore (and possibly establish) the 
general convenience of CC within QMC methods.  We are presently 
experiencing on atoms with more shells than lithium and trying to 
incorporate the remainder of Umrigar's method into our scheme.

\section{\label{Conclusions} Conclusions and perspectives}
The atomic shell structure originates from the peculiar radial 
dependence of the electron-nuclear potential $-Z/r$ and from the Fermi 
statistics; its typical multiplicity of length scales is not 
dramatically altered by the electron-electron repulsion, and is 
thus intimately related to the electron-nuclear distance.
Starting from F. Gygi's ideas, we have proposed a CC transformation 
which can stretch electron-nuclear distances, works for valence 
properties of full-core LDA atoms, and seems to improve variational 
Monte Carlo simulations of full-core atoms.
The atomic results of this paper represent an encouraging starting 
point for further developments and applications, which are underway: 
on one hand, plane-wave and real-space LDA calculations of full-core 
crystals and molecules; on the other, more investigations in the field 
of quantum Monte Carlo.\\

\noindent{\bf Acknowledgements}\\

\noindent We are grateful to C. Filippi for her precious training and 
assistance in the initial part of our VMC work, and to F. Gygi, G. 
Galli, S. Moroni and C. Umrigar for very useful conversations.  GBB 
gratefully acknowledges partial financial support from the Italian 
National Research Council (CNR, Comitato per la Scienza e la 
Tecnologia dell'Informazione, grants no.  96.02045.CT12 and 
97.05081.CT12).\par

\end{document}